\documentclass[manuscript]{aastex}

\begin{document}

\title{Generation of High-Energy Photons at Ultra-Relativistic Shock Breakout in Supernovae}
\author{Yukari Ohtani\altaffilmark{1,3}, Akihiro Suzuki\altaffilmark{2} and Toshikazu Shigeyama\altaffilmark{3}}
\altaffiltext{1}{Department of Astronomy, Graduate School of Science, University of Tokyo, Bunkyo-ku, Tokyo 113-0033, Japan}
\altaffiltext{2}{Center for Computational Astrophysics, National Astronomical Observatory of Japan, Mitaka, Tokyo 181-8588, Japan}
\altaffiltext{3}{Research Center for the Early Universe, Graduate School of Science, University of Tokyo, Bunkyo-ku, Tokyo 113-0033, Japan}

\begin{abstract}
We present theoretical expectations for non-thermal emission due to the bulk Comptonization  at the ultra-relativistic shock breakout. We calculate the transfer of photons emitted from the shocked matter with a Monte Carlo code fully taking into account special relativity. As a hydrodynamical model, we use a self-similar solution of \citet{2005ApJ...627..310N}.  Our calculations reveal that the spectral shape exhibits a double peak or a single peak  depending on the shock temperature at  the shock breakout. If it is significantly smaller than the rest energy of an electron, the spectrum has a double peak. We also display a few example of light curves, and estimate the total radiation energy. In comparison with observations of $\gamma$-ray bursts, a part of the higher energy component in  the spectra and the total energy can be reproduced by some parameter sets. Meanwhile, the lower energy counterpart in the Band function is not reproduced by our results and the duration time seems too short to represent the entire event of a $\gamma$-ray burst. Therefore the subsequent phase will constitute the lower energy part in the spectrum. 
\end{abstract}

\keywords{gamma-ray burst: general ---radiative transfer---relativistic processes --- scattering --- shock waves---stars: Wolf–Rayet}

\section{Introduction}
Some core-collapse supernovae radiate luminous UV or X-ray emission at the initial stage of explosions. This phenomenon, called shock breakout is caused by shock waves passing through the stellar surface. When the shock wave propagates deep inside the star, high energy photons generated in the shocked matter cannot diffuse out of the shock front and stay in the downstream of the shock, while they can escape to the upstream after the shock breakout when the shock wave approaches the stellar surface. Shock breakout is known for its very high luminosity ($\sim10^{46}\,{\rm erg\,s^{-1}}$). The duration is comparable to the light crossing time of the progenitor radius, up to several hours. This brevity has made it difficult to detect shock breakout emission until recently. SN 1987A is one of the examples exhibiting a precursory UV flash in the echo reflected by the circumstellar ring. Various models were investigated to explain the observed temporal behavior \citep{1991ApJ...380..575L,1988ApJ...331..377A}. \citet{1992ApJ...393..742E} studied shock breakout in blue supergiants and estimated the color temperature of UV emission in SN 1987A. These studies assumed that the shock breakout radiates thermal emission.

XRO 080109/SN 2008D is a shock breakout serendipitously detected by the Swift/XRT. \citet{2008Natur.453..469S} reported results of the observations as follows. Its peak luminosity was $L_{X,p}\approx6.1\times10^{43}\,{\rm erg\,s^{-1}}$ and the total energy was $E_{X}\approx2\times10^{46}\,{\rm erg}$. A power-law with the photon index $\Gamma\approx2.3$ is well fitted to the high-energy component in its spectrum, which indicates that XRO 080109 includes non-thermal emission. They argued that the bulk Comptonization, which had been proposed as a mechanism to produce non-thermal spectra \citep{2007ApJ...664.1026W}, may be responsible for such a power-law feature.  \citet{2010ApJ...719..881S} calculated transfer of photons traveling in spherical stars by a Monte Carlo simulation and reproduced the observed power-law spectrum if the shock velocity is $\geq0.3c$ at the shock breakout. Here the speed of light in vacuum is denoted by $c$. They used the self-similar solution by \citet{1960CPAM...13..353S} for describing the shock emergence from the stellar surface, in which the flow is assumed to be non-relativistic.

The flow associated with the shock emergence can be ultra-relativistic. For example, jets associated with $\gamma$-ray bursts (GRBs) are accelerated to ultra-relativistic speeds. The Lorentz factor $\Gamma$ would exceed 100. Though photospheric emission from relativistic shock breakouts has been investigated by some groups \citep[e.g.][]{2010ApJ...716..781K,2012ApJ...747...88N}, most of them do not calculate detailed spectral features. \citet{2013ApJ...767..139B} studied GRB emission from spherically symmetric outflows expanding at constant speeds ($\Gamma$=300, 500). They assumed that the photospheric radius is over $10^{12}$ cm, which describes emission after the shock passes the stellar surface. Furthermore, the accelerating shock plays an essential part in shock breakout and the shock is later decelerated by the circumstellar matter.

In this paper, we investigate effects of bulk Comptonization on emission from ultra-relativistic shock breakout. We use the self-similar solution \citep{2005ApJ...627..310N} to describe the propagation of ultra-relativistic shock waves  approaching the stellar surface. We have built a Monte Carlo code to treat the bulk Comptonization by fully taking into account relativistic effects and calculated spectra and light curves for several sets of parameters specifying the feature of stellar atmospheres and the strength of the shock waves.

We describe the hydrodynamical model in Section \ref{sec:formulation} and the radiative process using the Monte-Carlo method in Section \ref{sec:MCcode}. In Section \ref{sec:results} we present results and Section \ref{sec:conclusions} concludes the paper.

\section{Models}\label{sec:formulation}
To obtain spectra of photons emitted from the shock breakout, we calculate the transfer of photons emitted from the shocked matter. The matter is assumed to be composed of pure oxygen because GRBs are associated with massive type Ic SNe. The progenitor of SN Ic is thought to be a Wolf-Rayet star. Therefore we suppose that ultra-relativistic shock breakout occurs in a massive Wolf-Rayet (WC) star with the oxygen-rich envelope. Since we deal with events occurring in the vicinity of the stellar surface, we approximate the dynamical behavior of the shock breakout by  the self-similar solution \citep{2005ApJ...627..310N} that assumes the plane parallel geometry.

\subsection{Hydrodynamics}
Since WC star is thought to have the radiative envelope, the density distribution $\rho(x)$ can be expressed using the distance $x$ from the surface as,
\begin{eqnarray}
\rho(x) =
\left\{
\begin{array}{ll}
bx^{3}&x \geq 0, \\
0&x < 0, \\
\end{array}
\right.
\label{eq:ini_density}
\end{eqnarray}
where $b$ is a constant. Governing equations are given by \citet{2005ApJ...627..310N}.
\begin{eqnarray}
&&\frac{d}{dt}\left(p\gamma^{4}\right)=\gamma^{2}\frac{\partial p}{\partial t}, \label{eq:ordi_basic01}\\
&&\frac{d}{dt}\ln\left(p^{3}\gamma^{4}\right)=-4\frac{\partial \beta}{\partial x}, \label{eq:ordi_basic02}\\
&&\frac{d}{dt}\left(pn^{-4/3}\right)=0, \label{eq:ordi_basic03}
\end{eqnarray}
where $p$ is the pressure, $\gamma$  the bulk Lorentz factor, $\beta$  the velocity in units of $c$, and $n$ is the number density. The time $t$ is measured from the moment when the shock reaches the stellar surface defined by $x=0$. Note that the velocity and the time have negative values.

The boundary conditions at the shock front are given by the Taub relations \citep{1948PhRv...74..328T}, which are reduced to
\begin{eqnarray}
&&p_{2}=\frac{2}{3}\Gamma^{2}\rho_{1}c^2,\label{eq:rankine-hugoniot-p}\\
&&\rho_{2}\gamma_{2}=2\rho_{1}\Gamma^{2},\\
&&\gamma_{2}^{2}=\frac{\Gamma^{2}}{2},
\end{eqnarray}
in the ultra-relativistic limit \citep{1976PhFl...19.1130B}. The subscripts 1 and 2 denote the pre-shock and post-shock regions, respectively. The shock Lorentz factor $\Gamma$ is found to evolve with time in the form of
\begin{eqnarray}
\Gamma^{2}=A(-t)^{-3(2\sqrt{3}-3)},\label{eq:def_shockgamma}
\end{eqnarray}
where $A$ is a constant \citep{2005ApJ...627..310N}. The gas temperature could be derived from equation (\ref{eq:rankine-hugoniot-p})
\begin{eqnarray}
T_{\rm s}=\left(\frac{3p_{2}}{a}\right)^{1/4}=\left(\frac{2bX_{\rm s}^{3}\Gamma^{2}}{a}\right)^{1/4},\label{eq:temperature}
\end{eqnarray}
by assuming radiation dominated matter. Here $a$ is the radiation constant and $X_{\rm s}$ is the shock position from the surface.

This hydrodynamical model has four parameters: the  location $X_{\rm i}$  and the Lorentz factor $\Gamma_{\rm i}$ of the shock front,  the optical depth $\tau_{\rm i}$ to the shock front, and $b$ in equation (\ref{eq:ini_density}). The subscript i indicates values at the initial moment when the calculation starts. At the moment of the shock breakout, the shock velocity $V_{\rm s}$ is equal to the diffusion velocity of photons $v_{\rm diff}$, consequently
\begin{eqnarray}
\left|V_{\rm s}\right|\sim v_{\rm diff}=\frac{c}{\tau_{\rm i}}.
\end{eqnarray}
In the ultra-relativistic limit, $|V_{\rm s}|$ approaches to the speed of light. Thus the shock breakout would occur when the optical depth becomes unity.
Even before shock breakout, scattering may modify the spectrum. Therefore we begin calculations from the moment when the shock front reaches $\tau_{\rm i}=3$. We have confirmed that a larger $\tau_{\rm i}$ does not affect the results. 

These three parameters are not independent. The optical depth $\tau_{\rm i}$ is expressed in terms of $X_{\rm i}$ and $b$,
\begin{eqnarray}\label{eq:initialtau}
\tau_{\rm i}=\frac{\sigma_{\rm kl}}{\mu m_{\rm H}}\int_{0}^{X_{\rm i}}bx^{3}dx
=\frac{\sigma_{\rm kl}bX_{\rm i}^{4}}{4\mu m_{\rm H}},
\end{eqnarray}
where $\mu (=2)$ is the mean molecular weight in the Oxygen-rich envelope. We evaluate the Klein-Nishina cross section $\sigma_{\rm kl}$ for a photon with the energy equal to five times the average energy of the incident photons in the observer frame in order for most photons to scatter about 10 times before running out from the stellar surface. We will use $X_{\rm i}$ and $\Gamma_{\rm i}$ as free parameters to specify the shock breakout.

\begin{figure}
\epsscale{.50}
\plotone{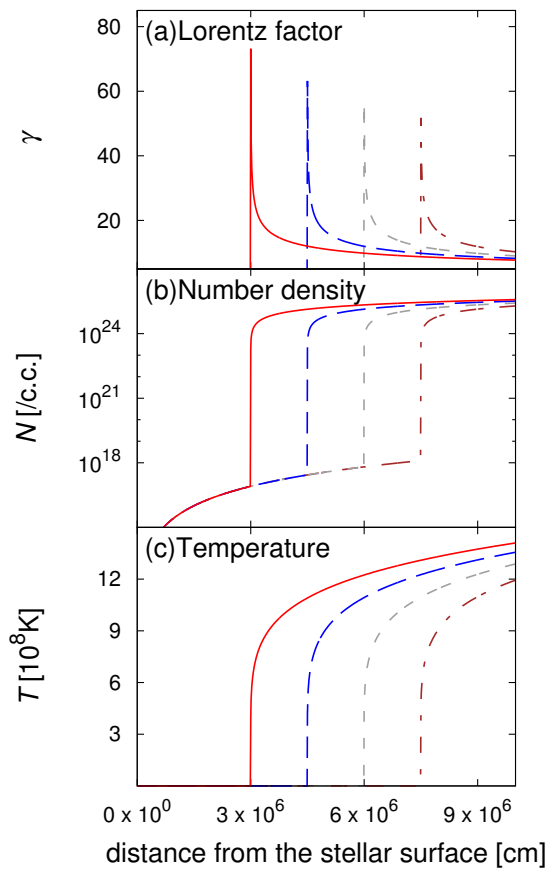}
\caption{The Lorentz factor, density, and temperature profile ($X_{\rm i}=10^{7}$ cm, $\Gamma=100$). Each time in the observer frame is $2.5\times10^{-4}$ s (dash-dotted line), $1.5\times10^{-4}$ s (short-dashed line), $2.0\times10^{-4}$ s (long-dashed line), and $1.0\times10^{-4}$ s (solid line).}
\label{fig:profile-Multi-g100gnt02-j}
\end{figure}

Figure \ref{fig:profile-Multi-g100gnt02-j} shows profiles of the Lorentz factor, density, and temperature around the shock front for $X_{\rm i}=10^{7}$ cm and $\Gamma_{\rm i}=100$.

\subsection{Absorption Process}
Before discussing the transfer of photons via Compton scattering during the shock breakout, we evaluate the effective optical thickness in the Oxygen-rich envelope to estimate the influence introduced by ignoring absorption processes. The cross section for bound-free absorption is given by
\begin{eqnarray}
\sigma_{\rm bf}=\left(\frac{64\pi ng}{3\sqrt{3}Z^{2}}\right)\frac{e^{2}}{4\pi\varepsilon_{0}\hbar c}a_{0}^{2}\left(\frac{\nu_{n}}{\nu}\right)^{3},
\end{eqnarray}
where $h\nu=R_{\rm y}Z^{2}/n^{2}$ ($R_{\rm y}$: Rydberg constant) is the ionization energy for the initial state $n$, $g$ the bound-free Gaunt factor, $e$ the elementary charge, $\hbar$ the Dirac constant, $\varepsilon_{0}$ the vacuum permittivity, and $a_{0}$ is the Bohr radius \citep{1979rpa..book.....R}.  In our calculation, the photon energy in the rest frame of an electron in the upstream of the shock front is higher than 100 keV. According to the above formula, $\sigma_{\rm bf}$  $\approx10^{-25}\,{\rm cm}^{2}$ at $h\nu=100$ keV for H-like oxygen ion. This is comparable with the Klein-Nishina cross section at this energy, i.e., $\sigma_{\rm kl}\approx10^{-25}\,{\rm cm}^{2}$. In actual situations, the effective charge of oxygen ions must be smaller than 8 and the  energy of most photons must be greater than 100 keV, both of which reduces $\sigma_{\rm bf}$
much smaller than that for Compton scattering. Thus Compton scattering ionizes most of ions rather than bound-free transitions. As a result, the free-free transitions become the predominant absorption process. The free-free absorption coefficient is given by
\begin{eqnarray}
\alpha_{\nu}^{\rm ff}=3.7\times10^{8}T^{-1/2}Z^{2}n_{\rm e}n_{\rm i}\nu^{-3}(1-e^{-h\nu /k_{\rm B}T})\bar{g_{\rm ff}} \quad \rm cm^{-1},
\end{eqnarray}
where $k_{\rm B}$ is the Boltzmann constant and $\bar{g_{\rm ff}}$ is the gaunt factor \citep{1979rpa..book.....R}. From the hydrodynamical calculation in Section \ref{sec:formulation}, the number densities of electrons $n_{\rm e}$ and ions $n_{\rm i}$ are inferred to be $n_{\rm e}<8n_{\rm i}\lesssim10^{17}\,{\rm cm^{-3}}$. The matter temperature $T$ in the photospheric region of the progenitor star is about $10^{4}$ K. Then the absorption coefficient is estimated as $\alpha_{\nu}^{\rm ff}\lesssim1.7\times10^{-16}\,{\rm cm}^{-1}$. Therefore the effective optical depth becomes
\begin{eqnarray}
\tau_{*}=\sqrt{\alpha_{\nu}^{\rm ff}(\alpha_{\nu}^{\rm ff}+\bar{n}_{\rm e}\sigma_{\rm kl})}X_{\rm i}\lesssim10^{-12}X_{\rm i}.
\end{eqnarray}
In the following, we deal only with $X_{\rm i}\leq10^{11}$ cm, so that the possibility of absorption would be no more than 10\%, even for $h\nu=100$ keV. Thus we consider only the inverse Compton scattering.

\section{Monte Carlo Code}\label{sec:MCcode}
In this section, we describe the procedure of our Monte Carlo simulation to solve the radiative transfer equation. We determine whether scattering occurs in every time interval of $\Delta t=t_{\rm sh}/N$, where $t_{\rm sh}$ is defined as $t_{\rm sh}=X_{\rm i}/c$ and $N=1000$.
\subsection{Energy Distributions of Photons and Electrons}
Seed photons are produced on the shock front at every time step.
Their energy distribution follows the Planck function
\begin{eqnarray}
f_{\rm ph}=\frac{2}{c^{2}}\frac{\nu^{2}}{\exp(h\nu/k_{\rm B}T_{\rm s})-1},
\end{eqnarray}
in the rest frame of the bulk motion of the matter. The temperature $T_{\rm s}$ changes with time according to Equation (\ref{eq:temperature}).

The direction of motion of each photon is specified by the inclination angle $\theta$ and the azimuth angle $\phi$. They are specified by two random numbers to make photons distribute isotropically in the rest frame. We generate 100 photons at every time step. After the Lorentz transformation, most photons are emitted in the radial direction in the observer frame.

Electrons follow the Maxwell-Boltzmann distribution,
\begin{eqnarray}
f_{\rm e}\propto\exp\left[-\frac{p_{\rm e}^{2}}{2m_{\rm e}k_{\rm B}T_{\rm e}}\right],
\end{eqnarray}
where $p_{\rm e}$ is the momentum of an electron, $m_{\rm e}$  the electron mass, and $T_{\rm e}$ is the electron temperature. 

\subsection{Photon Traveling}
When a photon travels for the time interval $\Delta t$, the probability of electron scattering is described as
\begin{eqnarray}
P=1-\exp(-\tau),\label{eq:scat_prop}
\end{eqnarray}
where $\tau$ is the optical depth of the photon described by
\begin{eqnarray}
\tau=\sigma_{\rm kl}c\Delta t\tilde{n}_{\rm e}.\label{eq:opti_depth}
\end{eqnarray}
Here we generate a random number $R_{\rm sc}$ in the range of [0:1]. If $R_{\rm sc}<P$, then the photon is scattered at a distance of  
\begin{eqnarray}
l=-\frac{\ln(1-R_{\rm sc})}{\sigma_{\rm kl}\tilde{n}_{\rm e}},
\end{eqnarray}
from the starting point.  
We choose the scattering angle in the electron rest frame from another random number following the Klein-Nishina formula,
\begin{eqnarray}
\frac{d\sigma}{d\Omega}=\frac{1}{2}r_{0}^{2}f(\epsilon_{\rm i}^\prime,\Theta)^{2}
\left[f(\epsilon_{\rm i}^\prime,\Theta)+f(\epsilon_{\rm i}^\prime,\Theta)^{-1}-\sin^{2}\Theta\right],\\
f(\epsilon,\Theta)=\frac{1}{1+(1-\cos\Theta)\epsilon/m_{\rm e}c^2},
\end{eqnarray}
where $\Theta$ denotes the scattering angle.
Then, the photon energy changes from $\epsilon_{\rm i}^\prime$ to $\epsilon_{\rm f}^\prime$ as
\begin{eqnarray}
\epsilon^{\prime}_{\rm f}=\frac{\epsilon_{\rm i}^{\prime}}{1+(1-\cos\Theta){\epsilon_{\rm i}^{\prime}}/{m_{\rm e}c^{2}}}, \label{eq:inv_Compton_energy}
\end{eqnarray}
after scattering. When a photon propagating in the upstream is overtaken by the shock front,  we assume the photon immediately scatters off an electron at the shock front. 

After calculating changes of four vectors of photons according to the above procedures, we count the number of photons having escaped out of the stellar surface in each energy bin to construct a spectrum. Note that the number of seed photons should be normalized so that  the total number $\Delta N_{\nu}$ of the photons during the time interval of $\Delta t$ becomes
\begin{eqnarray}\label{eq:numberofphotons}
\Delta N_{\nu}=\frac{{4k_{\rm B}}^{3}T_{\rm s}^{3}\zeta(3)\Delta t}{h^{3}c^{2}\Gamma},
\end{eqnarray}
where $\zeta(3)=\sum^{\infty}_{n=1}1/n^{3}$ is the Riemann zeta function.

\section{Results}\label{sec:results}
We calculate spectra and light curves of  photons emitted from relativistic shock waves associated with jets. Here we assume that the angular aperture of the jet is 10 deg.
\subsection{Propagation of photons}
 We present an overall picture of how photons propagate in the shock breakout event, which summarizes results of our Monte Carlo simulations.
 
\begin{figure}[htb]
\epsscale{0.7}
\plotone{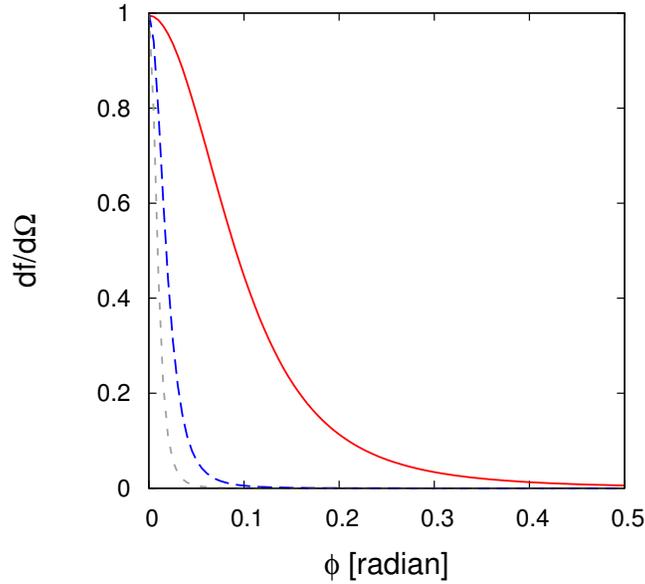}
\caption{The angular distribution of photons emitted from the shock front with the Lorentz factor of $\Gamma=$10 (red solid line), 50 (blue long-dashed line),100 (gray short-dashed line) in the observer frame.}
\label{fig:angular-distribution}
\end{figure}

Photons emitted from the shock front have an isotropic distribution in the rest frame of the shocked matter. Since the shocked matter has the Lorentz factor of $\Gamma/\sqrt{2}$, the angular distribution of photons in the observer frame has the form of $2/[\Gamma^2(1-\sqrt{1-2/\Gamma^2}\cos\phi)^2]$ where $\phi$ denotes the angle between the momentum of a photon and the shock normal. Figure \ref{fig:angular-distribution} shows the distributions for three shock Lorentz factors $\Gamma=$10, 50, 100. Photons with $\cos\phi>\sqrt{1-1/\Gamma^2}$ (about two thirds of total photons in the limit of large $\Gamma$) proceed ahead of the shock front. These photons scatter off electrons in the upstream of the shock. If the energy of a photon is smaller than or comparable to the rest energy ($m_{\rm e}c^2$) of an electron, i.e. $\Gamma k_{\rm B}T<m_{\rm e}c^2$, this scattering does not  appreciably change  the photon energy. If the energy of a photon is larger than $m_{\rm e}c^2$, then the scattering will reduce the energy to $\sim m_{\rm e}c^2$. After the scattering, most of the photons will be overtaken by the shock front and scatter off electrons in the shocked matter close to the front. This scattering increases the photon energy up to $\sim\Gamma(1+\cos\phi)m_{\rm e}c^2/\sqrt{2}$. 
If the optical depth of the upstream is still large, similar scattering events will repeat. Thus a photon can attain the maximum energy of $\sim\Gamma_{\rm ph}m_{\rm e}c^2$ where $\Gamma_{\rm ph}$  is the shock Lorentz factor when the front reaches the photosphere ($\tau=1$).

\subsection{Spectra}\label{sec:spectra}
\begin{figure}
\epsscale{1.0}
\plotone{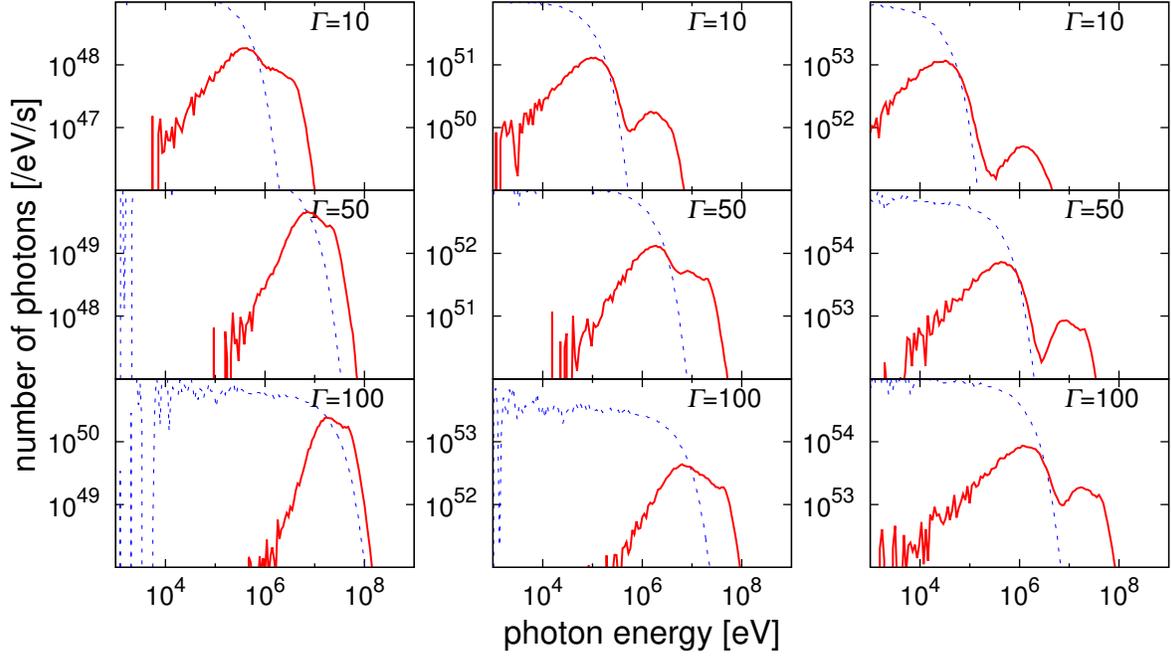}
\caption{Spectra. The free parameters are $(R,\,X_{\rm i})$=($10^{10}$ cm, $10^{7}$ cm) (left), ($10^{12}$ cm, $10^{9}$ cm) (center), ($10^{13}$ cm, $10^{11}$ cm) (right), and $\Gamma_{\rm i}$=10 (top), 50 (center), 100 (bottom), respectively. The spectral figures are unaffected by the stellar radius.}
\label{fig:Multi-k10-100t3x7-11dim-j}
\end{figure}

Figure \ref{fig:Multi-k10-100t3x7-11dim-j} shows spectra for 9 parameter sets of $X_{\rm i}=10^7\,{\rm cm},\, 10^9\,{\rm cm},\, 10^{11}\,{\rm cm}$ and  $\Gamma_{\rm i}$=10, 50, 100. Though the radius $R$  does not affect the spectral shape in the logarithmic scale, we have assumed that $R=10^{10}$ cm for $X_{\rm i}=10^7\,{\rm cm}$, $R=10^{12}$ cm for $X_{\rm i}=10^9\,{\rm cm}$, $R=10^{13}$ cm for $X_{\rm i}=10^{11}\,{\rm cm}$ to evaluate the absolute value of differential luminosity per unit photon energy. 

The broken line in each panel  exhibits spectrum of seed photons emitted from the shock front, which is a superposition of the Planck distributions with different temperatures. The spectral shape of escaping photons (solid line) shows a single peak when $X_{\rm i}=10^{7}$ cm. Since the peak is composed of photons generated at $\tau\lesssim1$, it locates at $E_{\alpha}\sim{\Gamma_{\rm ph} k_{\rm B}T_{\rm s,\, ph}}$. Here $T_{\rm s,\, ph}$ denotes the shock temperature at $\tau=1$. Meanwhile scattered photons coming from $\tau\sim\tau_{\rm i}$ gain energies up to $E_{\beta}\sim\Gamma_{\rm ph} m_{\rm e}c^{2}\propto\Gamma_{\rm i}$ and form a cut off above this energy in the spectra.

As $X_{\rm i}$ increases, the spectral shape turns into a double-peak structure. This is because the maximum energy $E_{\alpha}$ ($\propto T_{\rm s,\, ph}\propto X_{\rm s}^{3/4}/X_{\rm i}$) of seed photons  become significantly lower than the kinetic energy of an electron in the shocked matter. The large gap between these energies produces the double-peak structure. Most of photons scattered in the upstream of the shock front are necessarily scattered by the shocked matter and then increase the energy up to $\Gamma_{\rm ph} m_{\rm e}c^{2}$ while those directly escaping from the surface retains the energy $\sim E_{\alpha}$. Therefore photons scarcely have energies in between.
We describe the shape around the lower energy peak using two distinct power-law functions, of which the exponents are denoted by $\alpha$ and $\beta$. Table \ref{tb:sp-character01} lists several parameters characterizing the spectra, where $E_{\xi}$ is the energy of the flux minimum between the two peaks.
\begin{table}[htb]
\begin{center}
\begin{tabular}{|c|c||r|c|r|r|r|r|r|r|}
\hline
$X_{\rm i}$&$\Gamma_{\rm i}$&$k_{\rm B}T(\tau=3)$&Double Peak?&$\Gamma_{\rm ph}$&$E_{\alpha}$&$\alpha$&$E_{\beta}$&$\beta$&$E_{\xi}$\\ \hline
&10&$13\,{\rm keV}$&No&12&$0.4\,{\rm MeV}$&1.25&$4.0\,{\rm MeV}$&-0.57&-\\ \cline{2-10}
$10^{7}\,{\rm cm}$&50&$42\,{\rm keV}$&No&61&$6.3\,{\rm MeV}$&1.41&$22\,{\rm MeV}$&-0.67&-\\ \cline{2-10}
&100&$76\,{\rm keV}$&No&121&$16\,{\rm MeV}$&1.94&$48\,{\rm MeV}$&-0.60&-\\ \hline
&10&$3.4\,{\rm keV}$&Yes&12&$120\,{\rm keV}$&0.93&$1.6\,{\rm MeV}$&-2.20&$630\,{\rm keV}$\\ \cline{2-10}
$10^{9}\,{\rm cm}$&50&$11\,{\rm keV}$&Yes&61&$2.0\,{\rm MeV}$&1.20&$20\,{\rm MeV}$&-0.79&$5.5\,{\rm MeV}$\\ \cline{2-10}
&100&$18\,{\rm keV}$&No&121&$6.6\,{\rm MeV}$&0.82&$41\,{\rm MeV}$&-2.73&-\\ \hline
&10&$1.0\,{\rm keV}$&Yes&12&$34\,{\rm keV}$&0.74&$1.2\,{\rm MeV}$&-2.81&$270\,{\rm keV}$\\ \cline{2-10}
$10^{11}\,{\rm cm}$&50&$2.8\,{\rm keV}$&Yes&61&$500\,{\rm keV}$&0.64&$8.0\,{\rm MeV}$&-2.43&$2.8\,{\rm MeV}$\\ \cline{2-10}
&100&$4.7\,{\rm keV}$&Yes&121&$1.3\,{\rm MeV}$&0.58&$17\,{\rm MeV}$&-1.90&$7.2\,{\rm MeV}$\\
\hline
\end{tabular}
\end{center}
\caption{Comparison of our spectra with the Band function. $E_{\alpha}$ corresponds to the peak energy of the Band function between two power-law functions, of which the exponents are denoted by $\alpha$ and $\beta$. $E_{\beta}$ indicates the higher energy peak in a double-peak spectrum, while it is the cut off energy in a single-peak spectrum. The flux minimum of the double-peak locates at $E_{\xi}$.}
\label{tb:sp-character01}
\end{table}

We compare our results with typical GRB spectra, i.e. the Band function \citep{1993ApJ...413..281B}. The  peak energies in observed spectra are located below 1 MeV.  Figure \ref{fig:Multi-k10-100t3x7-11dim-j} suggests that this condition is satisfied by models with the initial Lorentz factor less than 50 and the initial position of the shock $X_{\rm i}$ larger than $10^{9}$cm. The power-law exponent $\beta$ of the higher energy component in the Band function is typically $\sim-2.5$. When  photons form a double peak spectrum, this condition seems to be satisfied. On the contrary, the exponent $\alpha$ of lower energy component in the Band function is $\sim-1$, which does not appear to match the lower energy components of our results, which have positive exponents. This lower energy component might be formed by emissions from the subsequent phases of the jet propagation. 

\subsection{Light Curves}
\begin{figure}[htb]
\epsscale{0.6}
\plotone{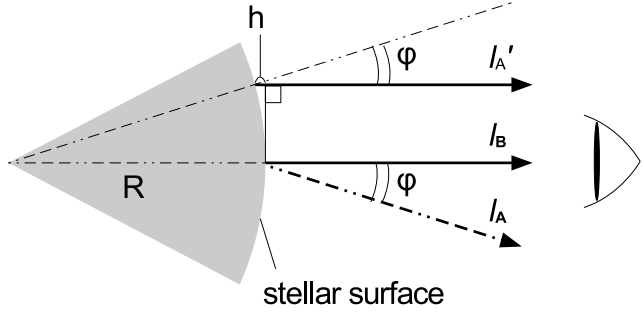}
\caption{The sectional drawing of the progenitor star. The jet angle is supposed to be $10$ deg.}
\label{fig:URSB-LCmake010402-20130117}
\end{figure}

\begin{figure}[htb]
\epsscale{1.0}
\plotone{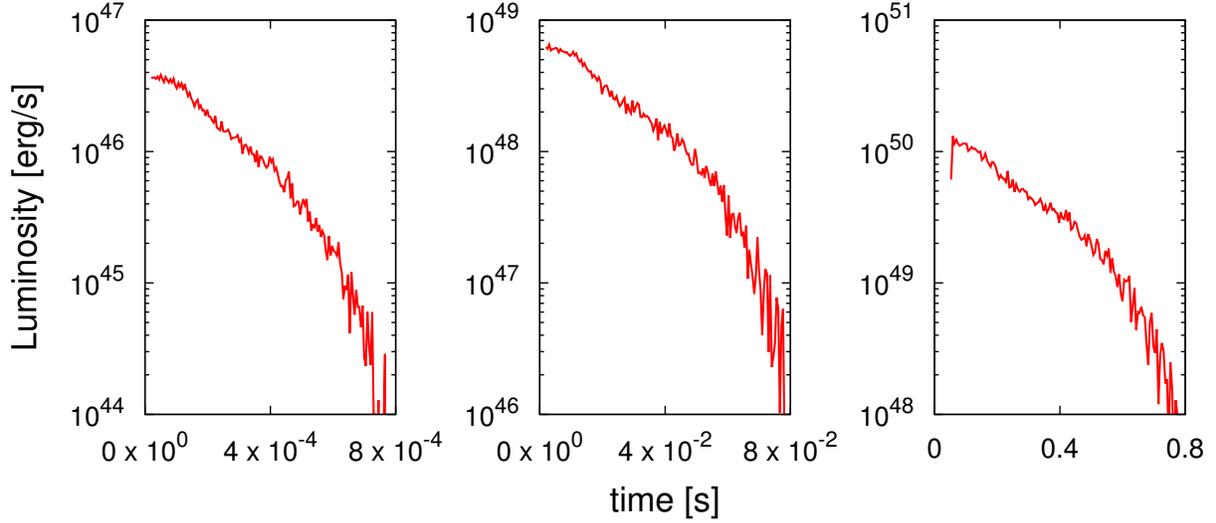}
\caption{Light Curves. (left panel: $R=10^{10}$ cm, $X_{\rm i}=10^{7}$ cm, $\Gamma_{\rm i}=10$, center panel : $R=10^{12}$ cm, $X_{\rm i}=10^{9}$ cm, $\Gamma_{\rm i}=10$, right panel: $R=10^{13}$ cm, $X_{\rm i}=10^{11}$ cm, $\Gamma_{\rm i}$=10.)}
\label{fig:tg10t3-Multi-r13-10dim-j}
\end{figure}

We calculate light curves from relativistic jets emanating from the photosphere by assuming axial symmetry. To determine the timing of detecting each photon, we take into account the difference of traveling time due to different directions of motion of photons.
Figure \ref{fig:URSB-LCmake010402-20130117} displays the path difference $h$ of photons going out of the surface in different directions. Because of the assumed axial symmetry, the photon propagating along the line $l_{\rm A}^{\prime}$ is regarded to be equivalent to that along $l_{\rm A}$.
The lag time between the emission $l_{\rm A}^{\prime}$ and $l_{\rm B}$ at the observer is expressed as
\begin{equation}
\delta t=\frac{h}{c}=\frac{R}{c}\left(1-\cos\varphi\right).
\end{equation}
The arrival time at the observer is given by $t_{\rm obs}=t_{\rm f}+\delta t$, where $t_{\rm f}$ is the time when the photon passes the stellar surface. We calculate light curves as functions of the observed time $t_{\rm obs}$. The results for three sets of parameters are shown in Figure \ref{fig:tg10t3-Multi-r13-10dim-j}. Each of them is well fitted with a power-law with exponent $-1.0\sim-0.9$ in the range of $0.1t^{99}_{\rm obs}<t<0.9t^{99}_{\rm obs}$, where $t^{99}_{\rm obs}$ is the time when 99\% of scattered photons have arrived at the observer ($t^{99}_{\rm obs}=5.9\times10^{-4}$ s for $(R,\,X_{\rm i},\,\Gamma_{\rm i})$=($10^{10}$ cm, $10^{7}$ cm, 10),  $6.1\times10^{-2}$ s for $(R,\,X_{\rm i},\,\Gamma_{\rm i})$=($10^{12}$ cm, $10^{9}$ cm 10), 0.6 s for $(R,\,X_{\rm i},\,\Gamma_{\rm i})$=($10^{13}$ cm, $10^{11}$ cm, 10), respectively). The duration becomes of the order of $R/c\Gamma_{\rm i}^{2}$.  It is clear that none of them reproduces the duration of a typical long GRB prompt emission longer than 2 s. The shock breakout occupies a tiny fraction of the duration of a GRB event. The subsequent activities of the central engine are responsible for the rest of the prompt emission and may constitute the emission of lower energies. 

We can derive the dependence of the energy of radiation emitted from the jet on the parameters $R$ and $\Gamma_{\rm i}$ as follows. From equation (\ref{eq:numberofphotons}),  the number of generated photons per unit area is shown to be
\begin{eqnarray}
N_{\nu}=\frac{32  \zeta (3)k_{\rm B}^3 }{\left(11-3 \sqrt{3}\right) c^{3/2} h^3}\left(\frac{3m_{\rm H} m_{\rm u}}{a \sigma_{\rm kl}}\right)^{3/4}\sqrt[4]{2X_{\rm i}}  \sqrt{\Gamma_{\rm i}},
\end{eqnarray}
using the  dependence of the temperature on $X_{\rm i}$ and $\Gamma_{\rm i}$ derived from equations (\ref{eq:temperature}) and (\ref{eq:initialtau}) for a given $\tau_{\rm i}(=3)$.
The energy of scattered photons is dependent on $X_{\rm i}$. The average energy of thermal photons emitted from the shock front is inversely proportional to $X_{\rm i}$. As a consequence, a larger $X_{\rm i}$ converts the energy of scattered  photons to $\Gamma_{\rm ph}k_{\rm B}T_{\rm s,\,ph}$ while a smaller $X_{\rm i}$ increases the energy up to $\sim\Gamma_{\rm ph}m_{\rm e}c^{2}$ as was discussed in section \ref{sec:spectra}.
Then the  average luminosity does not follow a simple scaling law in terms of $X_{\rm i}$. On the other hand, it can be shown that the luminosity is proportional to $R\Gamma_{\rm i}^{7/2}$ because the emission region is proportional to the square of $R$, the photon energy is proportional to $\Gamma_{\rm i}$, and the duration of the event scales as $R/(c\Gamma_{\rm i}^2)$.


\section{Conclusions}\label{sec:conclusions}
We investigate effects of the bulk Comptonization at the shock breakout in the ultra-relativistic limit using a Monte Carlo code developed by the authors. In the calculations, we have three  parameters: the progenitor radius $R$, the shock position $X_{\rm i}$ where the Comptonization starts to affect the spectra ($\tau=3$), and the corresponding shock Lorentz factor $\Gamma_{\rm i}$.

The spectral shape is affected by $X_{\rm i}$ and $\Gamma_{\rm i}$. It exhibits double peak for larger $X_{\rm i}$, and single peak for smaller $X_{\rm i}$. If we regard the lower  energy peak at the photon energy of $E_{\alpha}$ as the break energy of the Band function, the spectrum with the parameters of $X_{\rm i}=10^{11}$ cm and $\Gamma_{\rm i}=10$ reproduces typical values of $E_\alpha$ and $\beta$ exhibited by GRBs. In respect to the total energy, $R > 10^{13}$ cm is suggested to match the observed radiation energy of a GRB, because the total energy of our results is proportional to $R^{2}$. The duration of the present models is far shorter than the long-GRB duration ($\geq2$ sec). Therefore our models can be applied only to the very early stage of GRBs. Recent surveys by Fermi/LAT show that photons with energies above 100 MeV tend to delay by several seconds in the arrival time \citep{2009Natur.462..331A,2009Sci...323.1688A}. The result of our calculations that few photons attain energies higher than 100 MeV is consistent with the lack of these high energy photons in the very early phase. There remains a discrepancy between the spectra of models and observed GRB. The shape of the lower energy component in the Band function ($N(E)\propto E^{-1}$) is not reproduced by our results. There may be some possibility of reducing this discrepancy if we add some thermal radiation from the optically thick ejecta propagating in the circumstellar matter after the shock breakout.

In this study, we do not take into account the energy loss of the shocked matter through photon diffusion or the energy gain of the envelope in front of the shock by inverse Compton scattering. To study these effects, we need to perform radiation hydrodynamic calculations, which is beyond the scope of this paper. Furthermore, the structure of radiation mediated shocks has been investigated \citep[e.g.][]{2010ApJ...716..781K,2010ApJ...725...63B}. If an electron-positron plasma layer is produced immediately behind the shock, the photospheric emission  should be affected by inverse-Compton scattering in this region.


\bibliographystyle{apj}
\bibliography{hogehoge}

\end{document}